\documentclass[12pt]{article}  %modified 12/16/06, 9/2/07, 9/3/16, 3/9/19 csb
\newcommand{\ben}{\begin{enumerate}}
\newcommand{\bthm}{\begin{thm}}
\newcommand{\bdfn}{\begin{dfn}}
\newcommand{\ba}{\begin{array}}
\newcommand{\be}{\begin{equation}}
\newcommand{\bea}{\begin{eqnarray}}
\newcommand{\ethm}{\end{thm}}
\newcommand{\edfn}{\end{dfn}}
\newcommand{\ea}{\end{array}}
\newcommand{\ee}{\end{equation}}
\newcommand{\eea}{\end{eqnarray}}
\newcommand{\een}{\end{enumerate}}

\begin{document}
\bibliographystyle{unsrt}

\title{\bf Prony, Pad\'e, and Linear Prediction for Interpolation and Approximation
    in the  \\ Time and Frequency Domain Design of IIR Digital Filters and in Parameter Identification}

\author{{\it C. Sidney Burrus }\\
    Electrical and Computer Engineering Department \\
    and the OpenStax Project \\
    Rice University, Houston, Tx 77005-1892 \\
    csb@rice.edu\\
    Open license under Creative Commons (cc-by-4.0)
}

\maketitle

\section{Introduction}

Model based signal processing or signal analysis or signal representation
has a rather different point of view from the more traditional filtering
and algorithm based approaches.  However, in all of these, the names of Prony,
Pad\'e, and linear prediction come up.  This note examines these ideas
with the goal of showing they are all based on the same principles and all
can be extended and generalized.

Prony originally posed the problem of separating the individual
characteristics of gases from measurements made on a combination of them
 in 1795 \cite{prony-95}.
We pose this problem using modern terminology and notation as follows.
A signal is modeled as $y(n)$ which consists of equally spaced
samples of $x(t)$ which is a finite sum of $N$ exponentials,
\be
    x(t) = \sum_{k=1}^N \, K_k \, e^{\alpha_k t}.
\label{p0}\ee
The sampling is described by
\be
    y(n) = x(Tn) = \sum_{k=1}^N \, K_k \, e^{\alpha_k T n} =
        \sum_{k=1}^N \, K_k \, \lambda_k^n
\label{p1}\ee
where $\lambda_k = e^{\alpha_kT}$.
Prony's problem is to calculate the $2N$ parameters, $K_k$ and $\alpha_k$
(or $\lambda_k$), from the samples, $y(n)$.

Prony's original approach \cite{prony-95} to solving this problem
is straight-forward but awkward, requiring two different procedures for
even order and odd order systems.
A more flexible and general method can be derived using the z-transform of
Prony's formulation.

Pad\'e's method \cite{pade-1892} is a rational approximation of a polynomial.
It has been shown \cite{weiss-mcdonough-63} to be the z-transformed
version of Prony's method.
The z-transform of (\ref{p1}) can be written as an $N^{th}$ order rational
function of $z$
\be
    Y(z) = \frac{b_0 + b_1 z^{-1} + \cdots + b_M z^{-M}}
    {1 + a_1 z^{-1} + \cdots + a_N z^{-N}}
\label{p2}\ee with $M = N-1$.    The problem is to find the $2N$
parameters, $b_k$ and $a_k$, from the appropriate samples $y(n)$. This can
be expressed as a matrix convolution and solved \cite{bp2,pb}. The
equivalence of Prony's method and Pad\'e's method is discussed in
\cite{weiss-mcdonough-63}. Soewito made use of this equivalence
\cite{soewito} as did Vargas \cite{irlsvb}.

Linear prediction uses a model of the signal which assumes that a particular
value of the signal can be expressed as a weighted linear combination of
the $N$ previous values of the signal \cite{markel,cadzow}.  Finding those
weights turns out to be exactly the same problem as finding the
denominator coefficients in (\ref{p2}) and equivalent to finding the
$\alpha_k$ or $\lambda_k$ in (\ref{p1}).

These methods can all be extended from an interpolation problem to a least
equation error method by using more samples $y(n)$ than the number of
unknown coefficients, $a_n$ and $b_n$ or $K_k$ and $\alpha_k$.
This is discussed in \cite{bp2,pb}.  Iterative methods have been developed
which solve an iterative reweighted least equation error to achieve a
least solution error result \cite{steiglitz-mcbride-65,CKS,shaw}.  McClellan has
shown this to be equivalent to a well known adaptive filtering method
\cite{lee}.

The basic idea behind these three methods is based on the elimination or
minimization of the {\it equation error}, which is a linear problem, rather
than the usual minimization of the {\it solution error}, which is a
nonlinear problem. Some of these formulations also uncouple the calculation
of the numerator and the denominator coefficients \cite{shaw,bp2}

It is possible to pose the problem in the frequency domain and also
minimize the equation error \cite{levy,pb}.  One advantage to doing
that is better conditioning from the uniform sampling of the frequency
response between zero and $\pi$ rather than using the $K$ samples at the
beginning of an infinite time interval \cite{pb,kb,smith,csbfsi} and to deal
with the case where the design criteria are given in the frequency domain.

Prony's method (or Pad\'e's method or linear prediction) in both the time
and frequency domain is not only important for its own sake, but also as a
method to start  iterative methods or being the base of 
iterative methods that minimize other error criteria.  There is much still
to be discovered about this important, interesting, and powerful set of
methods.

\subsection*{Literature on Prony's Method and Extensions}

Original \cite{prony-95,pade-1892}, Linear Algebra \cite{strang7}\\
Basic \cite{hildebrand-56,tuttle-71,weiss-mcdonough-63,bp2,handscomb-66,cuyt,burstein,
bultheel,cheneyp}\\
Related \cite{rnm,mcdonough-huggins-68,kikkawa-73,brophy-salazar-74,Jain-71}
\cite{schulz-68,miller-73,maria-fahmy-73,barrodale,blaricum,svensson-73}\\
Applications\cite{markel-72,shanks-67,bpw,brophy-salazar-73,mb3,kalman-58,blaricum-mittra-75,pb2,bpw}\\
Iteration \cite{yengst,steiglitz-mcbride-65,mcbride2,evans-fischl-73,chao-lu-73,lee,fan,stonick4}\\
\\
\section{The Exact Solution or the Interpolation Problem in the Time Domain (Classical Prony)}

The formulation of the time-domain IIR filter design problem and the
methods for its solution are a version of Prony's (actually Pad{\'e}'s)
method.

The z-transform transfer function for an IIR filter \cite{pb} is given by
\be
    H(z) = \frac{B(z)}{A(z)} = \frac{b_0 + b_1z^{-1} + \cdots + b_Mz^{-M}}
                    {1 + a_1z^{-1} + \cdots + a_Nz^{-N}}
        = h_0 + h_1z^{-1} + h_2z^{-2} + \cdots
\label{p3}\ee
In the time domain, this becomes a convolution
\be
    \sum^{N}_{n=0} a_n \ h_{i-n} = b_i
\label{p4}\ee
where $a_0=1$ and $i=0, 1, \cdots, M$
which can be expressed in matrix form by
\be
    \left[ \ba{c}
    b_0\\
    b_1\\
    \vdots\\
    b_M\\
    0\\
    \vdots\\
    0
    \ea \right]
    =
        \left[ \ba{ccccc}
        h_0 & 0 & 0 & \cdots & 0\\
        h_1 & h_0 & 0  &  & \\
        h_2 & h_1 & h_0 &  & \\
        \vdots & & & & \vdots \\
        h_L & & \cdots & & h_0
        \ea \right]
            \left[ \ba{c}
            1\\
            a_1\\
            \vdots\\
            a_N\\
            0\\
            \vdots\\
            0
        \ea \right]
\label{p5}\ee

Note that the $h_n$  in (\ref{p5}) are the infinitely enduring causal impulse response values of the IIR
filter, not the aliased version of it used in \cite{csbfsi} and (\ref{cc104}).
A more compact matrix notation is
\be
    \left[ \ba{c}
    {\bf b}\\
    {\bf 0}
    \ea \right] = \left[ \ba{c} {\bf H} \ea \right]
        \left[ \ba{c}
        {\bf a}\\
        {\bf 0}
        \ea \right]
\label{p6}\ee
where ${\bf H}$ is $(L+1)$ by $(L+1)$, ${\bf b}$ is length-$(M+1)$, and
${\bf a}$ is length-$(N+1)$.  Because the lower $L-N$ terms of the
right-hand vector of (\ref{p5}) are zero, the ${\bf H}$ matrix can be reduced by
deleting the right-most $L-N$ columns to give ${\bf H}_0$ which causes
(\ref{p6}) to become
\be
    \left[ \ba{c}
    {\bf b}\\
    {\bf 0}
    \ea \right] = \left[ \ba{c} {\bf H}_0 \ea \right]
        \left[ \ba{c}
        {\bf a}
        \ea \right]
\label{p7}\ee
Because the first element of ${\bf a}$ is unity, it is partitioned to remove the
unity term and the remaining length-$N$ vector is denoted  ${\bf a^*}$. The
simultaneous equations represented by (\ref{p7}) are uncoupled by further
partitioning of the ${\bf H}$ matrix as shown in
\be
    \left[ \ba{c}
    {\bf b}\\
    {\bf 0}
    \ea \right] = \left[ \ba{c} {\bf H}_1\\ {\bf h}_1 \ \ {\bf H}_2 \ea \right]
        \left[ \ba{c}
        1\\
        {\bf a^*}
        \ea \right]
\label{p8}\ee
where ${\bf H_1}$ is $(M+1)$ by $(N+1)$, ${\bf h}_1$ is length-$(L-M)$,
and ${\bf H}_2$ is $(L-M)$ by $N$. The lower $(L-M)$ equations are written
\be
    {\bf 0}  =  {\bf h}_1  +  {\bf H}_2 {\bf a}^*
\label{p9}\ee
or
\be
    {\bf h}_1  =  - {\bf H}_2 {\bf a}^*
\label{p10}\ee
which must be solved for ${\bf a}^*$. The upper $M+1$ equations of (\ref{p8}) are
written
\be
    {\bf b}   =   {\bf H}_1 {\bf a}
\label{p11}\ee
which allows the calculation of ${\bf b}$.

If $L = N + M$, ${\bf H}_2$ is square. If ${\bf H}_2$ is nonsingular, (\ref{p10})
can be solved exactly for the denominator coefficients in ${\bf a}^*$,
which are augmented by the unity term to give ${\bf a}$.  From (\ref{p11}), the
numerator coefficients in ${\bf b}$ are found.  If ${\bf H}_2$ is singular
\cite{ls} and there are multiple solutions, a lower order problem
can be posed.  If there are no solutions, the methods of the next section
can be used and/or the assumed order increased.

Note that {\it any} trade-off in the order of numerator and denominator can be prescribed.
If the filter is in fact an FIR filter, ${\bf a}$ is unity and ${\bf a}^*$ does
not exist. Under these conditions, (\ref{p11}) states that $b_n = h_n$, which is
one of the cases of FIR frequency sampling covered in Section 3.1 of
\cite{pb}.  Also note that there is no control over the stability of the
filter designed by this method.

The IIR digital filter implemented using the denominator coefficients found in
(\ref{p10}) and the numerator coefficients found in (\ref{p11}) will have an impulse response
$h(n)$ that exactly matches the desired values given in $h(n)$.  The designed filter will
have an impulse response that interpolates the first $L+1$ terms of the given $h(n)$ but
says nothing about the response after that.  This is true if the total number of unknown filter
coefficients, $a(k)$ and $b(k)$ are equal to the number of given values $h(n)$ of the
desired impulse response.

\section{An Approximate Solution: the Least Equation Error Problem in the Time Domain }

In order to obtain better practical filter designs or parameter identification, the interpolation
scheme of the previous section is extended to give an approximation design
method \cite{pb}. It should be noted at the outset that the method developed
in this section minimizes an equation-error measure and not the usual
solution or signal error measure.

The number of samples specified, $L+1$, will be made larger than
the number of filter coefficients, $M+N+1$. This means that ${\bf H}_2$ is
rectangular and, therefore, (\ref{p6}) cannot in general be satisfied. To
formulate an approximation problem, a length-$(L+1)$ error vector
${\bf \varepsilon}$ is introduced in (\ref{p6}) and (\ref{p7}) to give
\be
    \left[ \ba{c}
    {\bf b}\\
    {\bf 0}
    \ea \right] = \left[ \ba{c} {\bf H}_0 \ea \right]
        \left[ \ba{c}
        {\bf a}
        \ea \right] + \left[ {\bf \varepsilon} \right]
\label{p12}\ee
Equation (\ref{p10}) becomes
\be
    {\bf h}_1 - {\bf \varepsilon}  =  - {\bf H}_2 {\bf a}^*
\label{p12a}\ee
where now  ${\bf H}_2$  is rectangular with $(L-M) > N$. Using the same
methods as used to derive (\ref{p10}), the error ${\bf \varepsilon}$ is minimized in a
least-squared error sense by the solution of the normal equations \cite{ls,csbvs1} which occurs
when the error is orthogonal to the impulse response.
\be
    {\bf H}_2^T {\bf h}_1 = - {\bf H}_2^T {\bf H}_2 {\bf a}^*
\label{p13}\ee
Alternatively, this equation can also be derived by taking the gradient of the squared error
and setting it equal to zero (necessary condition for a minimum).

If the equations are not singular, the optimal solution is
\be
     {\bf a}^* = - [{\bf H}_2^T {\bf H}_2]^{-1} {\bf H}_2^T {\bf h}_1.
\label{p14}\ee
which uses the so-called pseudo-inverse \cite{ls,csbvs}.
If the normal equations are singular, the pseudo-inverse
can be used to obtain a minimum norm or reduced order solution.

The numerator coefficients are found by the same techniques as before in
(\ref{p11})
\be
    {\bf b}   =   {\bf H}_1 {\bf a}
\label{p15}\ee
which results in the upper $M+1$ terms in ${\bf \varepsilon}$ being zero
and the total squared equation error ($L_2$ norm) being minimum.

As is true for least-squared-error optimal design of FIR filters, (\ref{p13}) is can be numerically
ill-conditioned and (\ref{p14}) should not be used directly to solve for ${\bf a}^*$.
Special algorithms such as those used by Matlab and LINPACK
\cite{linpack} should be employed.

\section{General Optimal Design of Zeros of Transfer Function}

Given the pole locations designed by minimizing the equation error from (\ref{p11}) 
or from any other source, the zero locations (numerator coefficients) can be separately
designed to interpolate certain points, to further minimize the equation error, 
or to minimize the the solution error. 

Reformulate (\ref{p13}) to have the form
\be
    \left[ \ba{c}
    b_0\\
    b_1\\
    \vdots\\
    b_M\\
    0\\
    \vdots\\
    0
    \ea \right]
    + [ e ]
    =
        \left[ \ba{ccccc}
        a_0 & 0 & 0 & \cdots & 0\\
        a_1 & a_0 & 0  &  & \\
        a_2 & a_1 & a_0 &  & \\
        \vdots & & & & \vdots \\
        a_{N-1} & & &  \\
        0 & & & & 0\\
        0 & & \cdots & & a_0      
        \ea \right]
            \left[ \ba{c}
            h_0\\
            h_1\\
            \vdots\\
            h_{K-1}\\
            0\\
            \vdots\\
            0
        \ea \right] = Ah
\label{p5a}\ee
where $A$ is a $K$ by $K$ lower triangular matrix which is therefore nonsingular.
The measure of the solution error is the difference between the actual and
desired impulse response:
\be
	e(n) = h_d(n) - h(n).
\label{p5b}\ee
This can be written as a $K$ by $1$ vector and becomes
\be
	e = h_d - h.
\label{p5c}\ee
giving
\be
    \left[ \ba{c}
    {\bf b}\\
    {\bf 0}
    \ea \right] = \left[ \ba{c} {\bf A} \ea \right]
        \left[ \ba{c}
        {\bf h_d}
        \ea \right] + \left[ {\bf \varepsilon} \right]
\label{p5d}\ee
In terms of the two measures of error, we have
\be
    \left[ \ba{c}
    {\bf b}\\
    {\bf 0}
    \ea \right] + \left[ {\bf \varepsilon} \right] = \left[ \ba{c} {\bf A} \ea \right]
        \left[ \ba{c}
        {\bf h_d}
        \ea \right] 
and
    \left[ \ba{c}
    {\bf b}\\
    {\bf 0}
    \ea \right] = \left[ \ba{c} {\bf A} \ea \right]
        \left[ \ba{c}
        {\bf h_d} + {\bf e}
        \ea \right] 
\label{p5e}\ee

If $a$ is found to minimize the equation error according to (\ref{p5}) or (\ref{p6}),
then a consistent $b$ can be found to minimize the equation error with (\ref{p7}).  
However, one might want to mix the criteria and find the $b$ that minimizes  
the solution error $||e||_2$.  From (\ref{p5e}) we have
\be
   h_d + e = 
     A^{-1}\left[ \ba{c}
    {\bf b}\\
    {\bf 0}
    \ea \right] = D\left[ \ba{c}
    {\bf b}\\
    {\bf 0}
    \ea \right]
\label{p5f}\ee
where $D = A^{-1}$.  Partitioning $D$ gives
\be
	H_d + e =  [{\bf D}_1 {\bf D}_2]\left[ \ba{c}
    {\bf b}\\
    {\bf 0}
    \ea \right] = D_1 {\bf b}
\label{p5g}\ee
with $D_1$ being $KxM$.

The numerator coefficients that minimize the solution error norm is the solution
to the following normal equations
\be
	D_1^TD_1 b = D_1^T h_d.
\label{p5h}\ee

This formulation can be generalized \cite{bp2,pb} to allow equality constraints and
a variety of other specifications.

\section{Prony's Method in the Frequency Domain gives the Frequency-Sampling Design of IIR Filters}

     In this section a frequency-sampling design method is
developed such that the frequency response of the IIR filter will interpolate or
pass through the given samples of a desired response. This development is
parallel to that used for Prony's method in the time domain.
Since a causal IIR filter cannot have linear phase, the sampled response must
contain both magnitude and phase. The extension of the frequency-sampling 
method to a LS-error approximation can be done as
for the FIR filter \cite{pb}. The method presented in this section uses a
criterion based on the equation error rather than the more common
error between the actual and desired frequency response.
Nevertheless, it is a useful noniterative design method. Finally,
a general discussion of iterative design methods for least-squares frequency
response error will be given.

     The method for calculating samples of the frequency response
of an IIR filter can be reversed to design
a filter much the same way it was for the FIR filter using frequency sampling \cite{pb}.
The z-transform transfer function for an IIR filter is given by

\be
    H(z) = \frac{B(z)}{A(z)} = \frac{b_0 + b_1z^{-1} + \cdots + b_Mz^{-M}}
                    {1 + a_1z^{-1} + \cdots + a_Nz^{-N}}.
\label{cc99}\ee
The frequency response of the filter is given by setting $z
=e^{-j\omega}$. Using the notation

\be
    H(\omega) = H(z)|_{z=e^{-j\omega}}.
\label{cc100}\ee
Equally-spaced samples of the frequency response are chosen so
that the number of samples is equal to the number of unknown
coefficients in (\ref{cc99}). These $L+1$ = $M+N+1$ samples of this
frequency response are given by

\be
    H_k = H(\omega_k) = H(\frac{2\pi k}{L+1})
\label{cc101}\ee
and can be calculated from the length-$(L+1)$) DFTs of the numerator
and denominator (padded with zeros to the proper length).

\be
    H_k = \frac{{\cal DFT}\{b_n\}}{{\cal DFT}\{a_n\}} =
        \frac{B_k}{A_k}
\label{cc102}\ee
where the indicated division is term-by-term division for each value
of $k$. Multiplication of both sides of (\ref{cc102}) by $A_k$ gives

\be
    B_k = H_k A_k
\label{cc103}\ee

If the length-$(L+1)$ inverse DFT of $H_k$ is denoted by the length-
$(L+1)$ sequence $h_n$, equation (\ref{cc103}) becomes cyclic convolution
which can be expressed in matrix form by

\be
    \left[ \ba{c}
    b_0\\
    b_1\\
    \vdots\\
    b_M\\
    0\\
    \vdots\\
    0
    \ea \right]
    =
        \left[ \ba{ccccc}
        h_0 & h_L & h_{L-1} & \cdots & h_1\\
        h_1 & h_0 & h_L    &  & \\
        h_2 & h_1 & h_0 &  & \\
        \vdots & & & & \vdots \\
        h_L & & \cdots & & h_0
        \ea \right]
            \left[ \ba{c}
            1\\
            a_1\\
            \vdots\\
            a_N\\
            0\\
            \vdots\\
            0
        \ea \right]
\label{cc104}\ee

Note that the  $h_n$  in (\ref{cc104}) are not the impulse response values
of the filter (they are an aliased version of it) as used in the FIR case or
in (\ref{p5}). Using the same approach as used for Prony's method in the time
domain, a more compact matrix notation is

\be
    \left[ \ba{c}
    {\bf b}\\
    {\bf 0}
    \ea \right] = \left[ \ba{c} {\bf H} \ea \right]
        \left[ \ba{c}
        {\bf a}\\
        {\bf 0}
        \ea \right]
\label{cc105}\ee
where ${\bf H}$ is $(L+1)$ by $(L+1)$, ${\bf b}$ is length-$(M+1)$,
and ${\bf a}$ is length-$(N+1)$.  Because the lower $L-N$ terms of
the right-hand vector of (\ref{cc104}) are zero, the ${\bf H}$ matrix can be
reduced by deleting the right-most $L-N$ columns to give ${\bf H}_0$
which causes (\ref{cc105}) to become

\be
    \left[ \ba{c}
    {\bf b}\\
    {\bf 0}
    \ea \right] = \left[ \ba{c} {\bf H}_0 \ea \right]
        \left[ \ba{c}
        {\bf a}
        \ea \right]
\label{cc106}\ee

Because the first element of ${\bf a}$ is unity, it is partitioned
to remove the unity term and the remaining length-$N$ vector is
denoted  ${\bf a^*}$. The simultaneous equations represented by (\ref{cc106})
are uncoupled by further partitioning of the ${\bf H}$ matrix as
shown in \be
    \left[ \ba{c}
    {\bf b}\\
    {\bf 0}
    \ea \right] = \left[ \ba{c} {\bf H}_1\\ {\bf h}_1 \ \ {\bf H}_2 \ea \right]
        \left[ \ba{c}
        1\\
        {\bf a^*}
        \ea \right]
\label{cc107}\ee
where ${\bf H_1}$ is $(M+1)$ by $(N+1)$, ${\bf h}_1$ is
length-$(L-M)$, and ${\bf H}_2$ is $(L-M)$ by $N$. The lower $(L-M)$
equations are written

\be
    {\bf 0}  =  {\bf h}_1  +  {\bf H}_2 {\bf a}^*
\label{cc108}\ee
or
\be
    {\bf h}_1  =  - {\bf H}_2 {\bf a}^*
\label{cc109}\ee
which must be solved for ${\bf a}^*$. The upper $M+1$ equations of
(10) are written
\be
    {\bf b}   =   {\bf H}_1 {\bf a}
\label{cc110}\ee
which allows the calculation of ${\bf b}$.

If $L = N + M$, ${\bf H}_2$ is square. If ${\bf H}_2$ is
nonsingular, (\ref{cc109}) can be solved exactly for the denominator
coefficients in ${\bf a}^*$, which are augmented by the unity term
to give ${\bf a}$.  From (\ref{cc110}), the numerator coefficients in ${\bf
b}$ are found.  If ${\bf H}_2$ is singular \cite{ls,csbvs} and there
are multiple solutions, a lower order problem can be posed.  If
there are no solutions, the approximation methods must be
used and/or the assumed order increased.

Note that any order numerator and denominator can be prescribed. If
the filter is in fact an FIR filter, ${\bf a}$ is unity and ${\bf
a}^*$ does not exist. Under these conditions, (\ref{cc110}) states that $b_n
= h_n$, which is one of the cases of FIR frequency sampling covered
\cite{pb}.  Also note that when there is a non-trivial denominator, there is no control
over the stability of the filter designed by this method.

This approach uses of the DFT therefore does not allow the
possibility of unequally spaced frequency samples as was possible
for FIR filter design.

The frequency-sampling design of IIR filters is somewhat more
complicated than for FIR filters because of the requirement that
${\bf H}_2$ be nonsingular. As for the FIR filter, the samples of
the desired frequency response must satisfy the conditions to insure
that $h_n$ are real. The power of this method is its ability to
interpolate arbitrary magnitude and phase specification. In contrast
to most direct IIR design methods, this method does not require any
iterative optimization with the accompanying convergence problems.

As with the FIR version, because this design approach is an
interpolation method rather than an approximation method, the
results may be poor between the interpolation points. This usually
happens when the desired frequency-response samples are not
compatable  with what an IIR filter can achieve.  One solution to
this problem is the same as for the FIR case
\cite{pb}, the use of more frequency samples than the number of
filter coefficients and the definition of an approximation error
function that can be minimized. Another solution is choose another
desired frequency that is closer to what can be achieved.
There is also no simple restriction that
will guarantee stable filters.  If the frequency-response samples
are consistent with an unstable filter, that is what will be
designed.

A complication for the this approach to the design of IIR filters is the need to specify
both the magnitude and phase of the desired frequency response.
This is not a problem for the time-domain design of IIR filters or
for the frequency-domain design of linear phase FIR filters, but
for the frequency domain design of IIR filters, samples of the
desired frequency response are complex numbers which means
both the magnitude and the phase must be specified.

\section{Discrete Least-Squared Equation-Error IIR Filter Design in the Frequency Domain}

Using the same approach as used for the time-domain, the
interpolation scheme is extended to give an
approximation design method in the frequency domain \cite{pb,isccsp08}. 
It should again be noted at the
outset that the method developed in this section minimizes an
equation-error measure and not the usual frequency-response error
measure.

The number of frequency samples specified, $L+1$, is made
larger than the number of filter coefficients, $M+N+1$. This means
that ${\bf H}_2$ is rectangular and, therefore, (\ref{cc109}) cannot in
general be satisfied. To formulate an approximation problem, a
length-$(L+1)$ error vector ${\bf \varepsilon}$ is introduced in (\ref{cc106})
and (\ref{cc107}) to give

\be
    \left[ \ba{c}
    {\bf b}\\
    {\bf 0}
    \ea \right] = \left[ \ba{c} {\bf H}_0 \ea \right]
        \left[ \ba{c}
        {\bf a}
        \ea \right] + \left[ {\bf \varepsilon} \right]
\label{cc111}\ee
Equation (\ref{cc109}) becomes \be
    {\bf h}_1 - {\bf \varepsilon}  =  - {\bf H}_2 {\bf a}^*
\label{cc112}\ee where now  ${\bf H}_2$  is rectangular with $(L-M) > N$. Using
the same methods as used to derive (\ref{cc109}), the error ${\bf
\varepsilon}$ is minimized in a least-squared error sense by the
solution of the normal equations \cite{ls} \be
    {\bf H}_2^T {\bf h}_1 = - {\bf H}_2^T {\bf H}_2 {\bf a}^*
\label{cc113}\ee

If the equations are not singular, the solution is \be
     {\bf a}^* = - [{\bf H}_2^T {\bf H}_2]^{-1} {\bf H}_2^T {\bf h}_1.
\label{cc114}\ee If the normal equations are singular, the pseudo-inverse
\cite{ls,csbvs} can be used to obtain a minimum norm or reduced
order solution.

The numerator coefficients are found by the same techniques as
before in (\ref{cc110}) \be
    {\bf b}   =   {\bf H}_1 {\bf a}
\label{cc115}\ee
which results in the upper $M+1$ terms in ${\bf \varepsilon}$
being zero and the total squared equation error being minimum.

As is true for the least squared error design of FIR filters, (\ref{cc113}) is often
numerically ill-conditioned and (\ref{cc114}) should not be used to solve for
${\bf a}^*$. Special algorithms such as those used by Matlab and
LINPACK \cite{ml,linpack} should be employed.

The error ${\bf \varepsilon}$ defined in (\ref{cc111}) can better be
understood by considering the frequency-domain formulation. Taking
the DFT of (\ref{cc111}) gives

\be
    B_k  =   H_k A_k   +  \varepsilon
\label{cc116}\ee
where $\varepsilon$ is the error in trying to satisfy (\ref{cc106}) when the
equations are over-specified.  This can be reformulated in terms of
${\cal E}$, the difference between the frequency response samples of
the designed filter and the desired response samples, by dividing
(\ref{cc106}) by $A_k$ to give

\be
    {\cal E}_k   =  \frac{B_k}{A_k} - H_k  = \frac{\varepsilon_k}{A_k}
\label{cc117}\ee
where ${\cal E}$ is the error in the solution of the approximation
problem, and $\varepsilon$ is the error in the equations defining
the problem. The usual statement of a frequency-domain approximation
problem is in terms of minimizing some measure of ${\cal E}$, but
that results in solving nonlinear equations. The design procedure
developed in this section minimizes the squared error $\varepsilon$,
thus only requiring the solution of linear equations.  There is an
important relation between these problems. Equation (\ref{cc117}) shows that
minimizing $\varepsilon$ is the same as minimizing ${\cal E}$
weighted by $A$.  However, $A$ is unknown until after the problem is
solved.

Although this is posed as a frequency-domain design method, the
method of solution for both the interpolation problem and the LS
equation-error problem is the same as the time-domain Prony's
method, discussed in Section 7.5 of reference \cite{pb}.

\section{General Optimal Design of Zeros of Transfer Function}

Given the pole locations designed by minimizing the equation error, the zero locations can be separately
designed to interpolate certain points, to further minimize the equation error, or to minimize the the 
solution error. This is similar to what was said earlier for the time-domain formulation.

\section{Coupled and Uncoupled Formulation of IIR Filter Design}

The formulation of both the time-domain and frequency-domain design
problems in (\ref{p14}) and (\ref{cc109}) uncouple the calculation
of the denominator coefficients (poles) from the calculation of the
numerator coefficients (zeros) but it assumes an equal spacing of the
frequency domain samples (the DFT us used) and the psuedo inverse
allows no error weighting.  Soewito \cite{soewito} uses a different
formulation which allows arbitrary location of the frequency samples
and weights in the minimization of the squared equation error, but
it couples the calculation of the numerator and denominator coefficients.

\section{Comments}
Numerous modifications and extensions can be made to this method. If
the desired frequency response is close to what can be achieved by
an IIR filter, this method will give a design approximately the same
as that of a true least-squared solution-error method. It can be
shown that $\varepsilon = 0 \leftrightarrow {\cal E} = 0$. In some
cases, improved results can be obtained by estimating $A_k$ and
using that as a weight on $\varepsilon$ to approximate minimizing
${\cal E}$. There are iterative methods based on solving (\ref{cc114}) and
(\ref{cc115}) to obtain values for $A_k$.  These values are used as weights
on $\varepsilon$ to solve for a new set of $A_k$  used as a new set
of weights to solve again for $A_k$ \cite{pb}.  The solution of
(\ref{cc114}) and (\ref{cc115}) is sometimes used to obtain starting values for other
iterative optimization algorithms that need good starting values for
convergence.

To illustrate this design method a sixth-order lowpass filter was
designed with 41 frequency samples to approximate.  The magnitude of
those less than 0.2 Hz is one and of those greater than 0.2 is zero.
The phase was experimentally adjusted to result in a good magnitude
response.  The design was performed with Program 9 in the appendix
of \cite{pb} and the frequency response is shown in Figure 7-33 of
\cite{pb}.

 In this section an LS-error approximation method was posed to
design IIR filters. By using an equation-error rather than a
solution-error criterion, a problem resulted that required only the
solution of simultaneous linear equations.

Like the FIR filter version, the IIR frequency sampling design
method and the LS equation-error extension can be used for complex
approximation and, therefore, can design with both magnitude and
phase specifications.

As noted earlier, if the desired frequency-response samples are close to what an IIR
filter of the specified order can achieve, this method will produce
a filter very close to what a true least-squared error method would.
However, when the specifications are not consistent with what can be
achieved and the approximating error is large, the results can be
very poor and in some cases, unstable.  It is often difficult
to set realistic phase response specifications. With this method, it
is even more important to have a design environment that will allow
easy trial-and-error procedure.

Other works on this problem are
\cite{K&I:90,K&I,garcia,lee,kb}. Other references can
be found in \cite{pb,K&I:90}.  The Matlab command {\tt
invfreqz()} which is an inverse to the {\tt freqz()} command gives a
similar or, perhaps, the same result as the method described in this
note but uses a different formulation \cite{levy,smith}. %{sptb}
\cite{brophy,brophy2,evans} A particularly interesting new
Matlab program is $iirlpnorm.m$ which allows the design of {IIR} 
filters with different degree numerator and denominators.  It also
allows use of an $L_p$ error minimization.  Unfortunately, it requires
both the zeros and poles to be inside the unit circle.  The poles of
the transfer function should be inside the unit circle to unsure
stability but allowing some zeros to be outside the unit circle (as
in the linear phase {FIR} filter) can improve the phase response.\cite{csbjt}
Soewito \cite{soewito} and Jackson\cite{jackson} formulate an 
iterative algorithm where the magnitue and phase are separately
updated to give an optimal magnitue approximation.

\section{Alternative Frequency Domain Formulation}

The use of the {DFT} to reformulate the time domain statement of Prony's 
method into the frequency domain uncouples the calculation of the $a_k$ 
and $b_k$ but requires equally spaced samples of the desired frequency
response.  An alternative formulation used by Soewito \cite{soewito} 
requires the calculation of the $a_k$ and $b_k$ together but allows
somewhat arbitrary frequency sample spacing.

\section{Iterative Algorithms using Prony's Methods}

\cite{stonick4,SidAhmed,vb1,icassp01v, irlsvb,soewito}

\section{Other Optimal IIR Filter Design Methods}

Martinez/Parks design \cite{martinez}, Jackson's improvement \cite{lbj2},
others
\cite{saramaki,alkhairy2,RGH,limlee,mclang7,mclang8,lang11,K&I:90,K&I,J&L}
\cite{S&A,cain,graham,crosara,berchin,steiglitz5,deczky5,cadzow5,dudgeon2,soewito}

\section{Summary}

This note has developed a time-domain and a frequency-domain method to
design an IIR digital filter that interpolates desired samples or that
gives an optimal, least squared equation error approximation. These
methods are directly related to Prony, Pad\'e, and linear prediction. In
addition, they can be used to obtain good starting values for
iterative algorithms or iterated themselves to obtain optimal
approximations with other criteria
\cite{levy,CKS,steiglitz-mcbride-65,pb,lbj3}. \baselineskip 2.5ex

\bibliography{book,burrus,burrus2,burrus3,art,filter} \baselineskip 4.5ex

\begin{thebibliography}{10}

\bibitem{prony-95}
Gaspard Clair Francois Marie~Riche Prony.
\newblock Eassi exp{\'e}rimental et analytique: Sur les lois de la
  dilatabilit{\'e} des fluides {\'e}lastiques et sur celles de la force
  expansive de la vapeur de l'eau et de la vapeur de l'alkool, a differentes
  temp{\'e}ratures.
\newblock {\em Journal de l'Ecole Polytechnique (Paris)}, 1(2):24--76, December
  1795.

\bibitem{pade-1892}
H.~E. Pad\'{e}.
\newblock Sur la representation approch\'{e}e d'une fonction par des fractions
  rationnelles.
\newblock {\em Ann. Sci. Ec. Norm. Sup. (Paris)}, 3(9):1--93, 1892.
\newblock Suppl.

\bibitem{weiss-mcdonough-63}
L.~Weiss and R.~N. McDonough.
\newblock Prony's method, z-transforms and {P}ad\'{e} approximations.
\newblock {\em SIAM Rev.}, 5(2):145--149, April 1963.

\bibitem{bp2}
C.~S. Burrus and T.~W. Parks.
\newblock Time domain design of recursive digital filters.
\newblock {\em IEEE Transactions on Audio and Electroacoustics},
  18(2):137--141, June 1970.
\newblock also in {IEEE} Press DSP Reprints, 1972.

\bibitem{pb}
T.~W. Parks and C.~S. Burrus.
\newblock {\em Digital Filter Design}.
\newblock John Wiley \& Sons, New York, 1987.
\newblock Now contained in OpenStax Filter Design book.

\bibitem{soewito}
Atmadji~W. Soewito.
\newblock {\em Least Square Digital Filter Design in the Frequency Domain}.
\newblock PhD thesis, Rice University, ECE Department, Houston, TX 77251,
  December 1990.
\newblock At: http://scholarship.rice.edu/handle/1911/16483.

\bibitem{irlsvb}
Ricardo Vargas and C.~Sidney Burrus.
\newblock Iterative design of $l_p$ digital filters.
\newblock {\em {arXiv}}, 2012.
\newblock arXiv:1207.4526v1 [cs.IT] July 19, 2012.

\bibitem{markel}
J.~D. Markel and A.~H. Grey.
\newblock {\em Linear Prediction of Speech}.
\newblock Springer-Verlag, Berlin Heidelberg, 1976.

\bibitem{cadzow}
James~A. Cadzow.
\newblock Signal processing via least squares error modeling.
\newblock {\em Signal Processing Magazine}, 7(4):12--31, October 1990.

\bibitem{steiglitz-mcbride-65}
K.~Steiglitz and L.~E. McBride.
\newblock A technique for the identification of linear systems.
\newblock {\em IEEE Transactions on Automatic Control}, AC-10:461--464, October
  1965.

\bibitem{CKS}
C.~K. Sanathanan and J.~Koerner.
\newblock Transfer function synthesis as a ratio of two complex polynomials.
\newblock {\em IEEE Transactions on Automatic Control}, 8:56--58, January 1963.

\bibitem{shaw}
Arnab~K. Shaw.
\newblock Optimal design of digital {IIR} filters by model-fitting frequency
  response data.
\newblock {\em IEEE Transactions on Circuits and Systems--II}, 42:702--710,
  November 1995.

\bibitem{lee}
J.~H. McClellan and D.~W. Lee.
\newblock Exact equivalence of the {Steiglitz--McBride} iteration and {IQML}.
\newblock {\em {IEEE} Transactions on Signal Processing}, 39(2):509--512,
  February 1991.

\bibitem{levy}
E.~C. Levy.
\newblock Complex-curve fitting.
\newblock {\em {IRE} Transactions on Automatic control}, AC-4(1):37--43, May
  1959.

\bibitem{kb}
R.~Kumaresan and C.~S. Burrus.
\newblock Fitting a pole-zero filter model to arbitrary frequency response
  samples.
\newblock In {\em Proceedings of the IEEE International Conference on
  Acoustics, Speech, and Signal Processing}, pages 1649--1652, IEEE ICASSP-91,
  Toronto, May 14--17 1991.

\bibitem{smith}
Julius~O. Smith.
\newblock {\em Techniques for Digital Filter Design and System Identification
  with Application to the Violin}.
\newblock PhD thesis, EE Dept., Stanford University, June 1983.
\newblock IIR filter design on page 50.

\bibitem{csbfsi}
C.~S. Burrus.
\newblock Frequency-sampling and the frequency domain {Prony} method for the
  design of {IIR} filters.
\newblock Unpublished notes, ECE Dept., Rice University, 1999.

\bibitem{strang7}
Gilbert Strang.
\newblock {\em Linear Algebra and Learning from Data}.
\newblock Wellesley Cambridge, Wellesley, 2019.
\newblock math.mit.edu/learningfromdata.

\bibitem{hildebrand-56}
F.~B. Hildebrand.
\newblock {\em Introduction to Numerical Analysis}.
\newblock McGraw-Hill, 1956.
\newblock pp. 318--386.

\bibitem{tuttle-71}
David~F. Tuttle.
\newblock On fluids, networks, and engineering education.
\newblock In R.E. Kalman and N.~DeClaris, editors, {\em Aspects of Network and
  System Theory}, pages 591--612. Holt, Rinehart and Winston, 1971.

\bibitem{handscomb-66}
D.~C. Handscomb.
\newblock {\em Methods of Numerical Approximation}.
\newblock Pergamon Press, 1966.
\newblock p. 135, 125.

\bibitem{cuyt}
Annie Cuyt and Luc Wuytack.
\newblock {\em Nonlinear Methods in Numerical Analysis}.
\newblock Elsevier, Amsterdam, 1987.
\newblock North--Holland Mathematics Studies: 136.

\bibitem{burstein}
Joseph Burstein.
\newblock {\em Approximation by Exponentials, Their Extensions, and
  Differential Equations}.
\newblock Metrics Press, Boston, 1997.

\bibitem{bultheel}
Adhemar Bultheel and Patrick Dewilde, editors.
\newblock {\em Rational Approximation in Systems Engineering}.
\newblock Birkh{\"a}user, 1983.
\newblock Sections on Pad{\'e} approximations.

\bibitem{cheneyp}
E.~W. Cheney.
\newblock {\em Introduction to Approximation Theory}.
\newblock McGraw-Hill, New York, 1966.
\newblock Section on Pad{\'e} Approximation, pp. 173--180.

\bibitem{rnm}
R.~N. McDonough.
\newblock Representation and analysis of signals, part {XV}, matched exponents
  for the representation of signals.
\newblock Technical report, April 1963.

\bibitem{mcdonough-huggins-68}
R.~N. McDonough and W.~H. Huggins.
\newblock Best least squares representation of signals by exponentials.
\newblock {\em IEEE Transactions on Automatic Control}, AC-13:408--412, August
  1968.

\bibitem{kikkawa-73}
S.~Kikkawa.
\newblock Time domain design of recursive filters for discrete inputs.
\newblock {\em Electronics and Communication in Japan}, 56-A(8):18--25, 1973.

\bibitem{brophy-salazar-74}
F.~Brophy and A.~C. Salazar.
\newblock Recursive digital filters synthesis in the time domain.
\newblock {\em IEEE Transactions on Acoustics, Speech, and Signal Processing},
  ASSP-22:45--55, February 1974.

\bibitem{Jain-71}
V.~K. Jain.
\newblock Representation of sequences.
\newblock {\em IEEE Transactions on Audio and Electroacoustics},
  AU-19:208--215, September 1971.

\bibitem{schulz-68}
E.~R. Schulz.
\newblock Estimation of pulse transfer function parameters by
  quasilinearization.
\newblock {\em IEEE Transactions on Automatic Control}, AC-13:424--426, August
  1968.

\bibitem{miller-73}
G.~Miller.
\newblock Least-squares rational z-transform approximation.
\newblock {\em J. of the Franklin Inst.}, 295:1--7, January 1973.

\bibitem{maria-fahmy-73}
G.~A. Maria and M.~M. Fahmy.
\newblock $l^{m}_{p}$ approximation by exponentials.
\newblock {\em IEEE Transactions on Circuit Theory}, pages 71--74, January
  1973.

\bibitem{barrodale}
I.~Barrodale and D.~D. Olesky.
\newblock Exponential approximation using {Prony's} method.
\newblock In C.~T.~H. Baker, editor, {\em The Numerical Solution of Nonlinear
  Problems}.

\bibitem{blaricum}
Michael~L. van Blaricum.
\newblock A review of {Prony's} method techniques for parameter estimation.
\newblock In {\em Air Force Statistical Estimation Workshop}, May 1978.

\bibitem{svensson-73}
T.~Svensson.
\newblock An approximation method for time domain synthesis of linear networks.
\newblock {\em IEEE Transactions on Circuit Theory}, pages 142--144, March
  1973.

\bibitem{markel-72}
J.~D. Markel.
\newblock Digital inverse filtering - a new tool for format trajectory
  estimation.
\newblock {\em IEEE Transactions on Audio and Electroacoustics},
  AU-20:129--137, June 1972.

\bibitem{shanks-67}
J.~L. Shanks.
\newblock Recursive filters for digital processing.
\newblock {\em Geophysics}, 32:33--51, February 1967.

\bibitem{bpw}
C.~S. Burrus, T.~W. Parks, and T.~B. Watt.
\newblock A digital parameter-identification technique applied to biological
  signals.
\newblock {\em IEEE Transactions Biomedical Engineering}, 18(1):35--37, January
  1971.

\bibitem{brophy-salazar-73}
F.~Brophy and A.~C. Salazar.
\newblock Considerations of the {P}ad\'{e} approximant technique in the
  synthesis of recursive digital filters.
\newblock {\em IEEE Transactions on Audio and Electroacoustics},
  AU-21:500--505, December 1973.

\bibitem{mb3}
S.~K. Mitra and C.~S. Burrus.
\newblock A simple efficient method for the analysis of structures of digital
  and analog systems.
\newblock {\em Archiv f{\"u}r Elektronik und {\"U}bertragungstechnik},
  31(1):33--36, January 1977.

\bibitem{kalman-58}
R.~E. Kalman.
\newblock Design of a self optimizing control system.
\newblock {\em Transactions of the ASME}, 80:468--478, February 1958.

\bibitem{blaricum-mittra-75}
M.~L. {Van Blaricum} and R.~Mittra.
\newblock A technique for extracting the poles and residues of a system
  directly from its transient response.
\newblock {\em IEEE Transactions on Antennas and Prop.}, AP-23:777--781,
  November 1975.

\bibitem{pb2}
T.~W. Parks and C.~S. Burrus.
\newblock Applications of {Prony's} method to parameter identification and
  digital filtering.
\newblock In {\em Proceedings of the Fifth Annual Princeton Conference on
  Information Sciences and Systems}, page 255, Princeton, NJ, March 1971.

\bibitem{yengst}
W.~C. Yengst.
\newblock Approximation by iterative methods.
\newblock {\em {IRE} Transactions on Circuit Theory}, CT-9(2):152--162, June
  1962.

\bibitem{mcbride2}
K.~E. McBride, H.~W. Schaefgen, and K.~Steiglitz.
\newblock Time-domain approximation by iterative methods.
\newblock {\em {IEEE} Transactions on Circuit Theory}, CT-13:381--387, December
  1966.

\bibitem{evans-fischl-73}
A.~G. Evans and R.~Fischl.
\newblock Optimal least squares time-domain synthesis of recursive digital
  filters.
\newblock {\em IEEE Transactions on Audio and Electroacoustics}, AU-20:61--65,
  February 1973.

\bibitem{chao-lu-73}
K.~S. Chao and K.~S. Lu.
\newblock On sequential refinement schemes for recursive digital filter design.
\newblock {\em IEEE Transactions on Circuit Theory}, pages 396--401, July 1973.

\bibitem{fan}
Hong Fan and Milo\u{s} Doroslova\u{c}ki.
\newblock On {``Global Convergence"} of {Steiglitz--McBride} adaptive
  algorithm.
\newblock {\em IEEE Transaction on Circuits and Systems--II: Analog and Digital
  Signal Processing}, 40(2):73--87, February 1993.

\bibitem{stonick4}
Virginia~L. Stonick and S.~T. Alexander.
\newblock A relationship between the recursive least squares update and
  homotopy continuation methods.
\newblock {\em IEEE Transactions on Signal Processing}, 39(2):530--532,
  Feburary 1991.

\bibitem{ls}
C.~L. Lawson and R.~J. Hanson.
\newblock {\em Solving Least Squares Problems}.
\newblock Prentice-Hall, Inglewood Cliffs, NJ, 1974.
\newblock Second edition by SIAM in 1987.

\bibitem{csbvs1}
C.~S. Burrus.
\newblock Vector space methods in signal and system theory.
\newblock Unpublished notes, ECE Dept., Rice University, Houston, TX, 1994.
\newblock expanded and now in OpenStax.

\bibitem{csbvs}
C.~S. Burrus.
\newblock Vector space methods in signal and system theory.
\newblock Unpublished notes, ECE Dept., Rice University, 1992.

\bibitem{linpack}
J.~J. Dongarra, J.~R. Bunch, C.~B. Moler, and G.~W. Stewart.
\newblock {\em LINPACK User's Guide}.
\newblock SIAM, Philadelphia, PA, 1979.

\bibitem{isccsp08}
C.~Sidney Burrus and Ricardo~A. Vargas.
\newblock The direct design of recursive or {IIR} digital filters.
\newblock In {\em Proceedings of International Symposium on Communications,
  Control, and Signal Processing}, ISCCSP-08, Malta, March 2008.

\bibitem{ml}
Cleve Moler, John Little, and Steve Bangert.
\newblock {\em Matlab User's Guide}.
\newblock The MathWorks, Inc., South Natick, MA, 1989.

\bibitem{K&I:90}
Takao Kobayashi and Satoshi Imai.
\newblock Design of {IIR} digital filters with arbitrary log magnitude function
  by {WLS} techniques.
\newblock {\em IEEE Trans. on ASSP}, 38(2):247--252, February 1990.

\bibitem{K&I}
Takao Kobayashi and Satoshi Imai.
\newblock Complex {Chebyshev} approximation for {IIR} digital filters in
  iterative {WLS} technique.
\newblock In {\em Proceedings of the ICASSP-90}, page 13.D3.12, Albuquerque,
  NM, April 1990.

\bibitem{garcia}
C.~B. Garcia and W.~I. Zangwill.
\newblock {\em Pathways to Solutions, Fixed Points, and Equilibria}.
\newblock Prentice-Hall, Englewood Cliffs, NJ, 1981.

\bibitem{brophy}
F.~Brophy and A.~C. Salazar.
\newblock Considerations of the {Pade} approximant technique in the synthesis
  of recursive digital filters.
\newblock {\em IEEE Transactions on Audio and Electroacoustics}, 21:500--505,
  1973.

\bibitem{brophy2}
F.~Brophy and A.~C. Salazar.
\newblock Recursive digital filter systhesis in the time domain.
\newblock {\em IEEE Transactions on Acoustics, Speech, and Signal Processing},
  22:45--55, 1974.

\bibitem{evans}
A.~G. Evans and R.~Fischl.
\newblock Optimal least squares time-domain systhesis of recursive digital
  filters.
\newblock {\em IEEE Transactions on Audio and Electroacoustics}, 21:61--65,
  1973.

\bibitem{csbjt}
Jasper Tan and C.~Sidney Burrus.
\newblock Near linear-phase {IIR} filters via {Gauss-Newton} method.
\newblock In {\em Proceedings of the Midwest Symposium on Circuits and
  Systems}, Dallas, Tx, August 4-7 2019.

\bibitem{jackson}
L.~B. Jackson.
\newblock {\em Digital Filters and Signal Processing}.
\newblock Kluwer Academic Publishers, Boston, MA, third edition, 1996.

\bibitem{SidAhmed}
M.~A. Sid-Ahmed, A.~Chottera, and G.~A. Jullien.
\newblock Computational techniques for least-square design of recursive digital
  filters.
\newblock {\em IEEE Transactions on Acoustics, Speech, and Signal Processing},
  26:478--480, 1978.

\bibitem{vb1}
Ricardo~A. Vargas and C.~Sidney Burrus.
\newblock Adaptive iterataive reweighted least squares design of $l_p$ filters.
\newblock In {\em Proceedings of the IEEE International Conference on
  Acoustics, Speech, and Signal Processing}, volume {III}, IEEE ICASSP-99,
  Phoenix, March 14--19 1999.

\bibitem{icassp01v}
Ricardo~A. Vargas and C.~Sidney Burrus.
\newblock On the design of {$L_p$ IIR} filters with arbitrary frequency
  response.
\newblock In {\em Proceedings of IEEE International Conference on Acoustics
  Speech and Signal Processing}, ICASSP-01, Salt Lake City, May 7-11 2001.

\bibitem{martinez}
H.~G. Martinez and T.~W. Parks.
\newblock Design of recursive digital filters with optimum magnitude and
  attenuation poles on the unit circle.
\newblock {\em IEEE Transactions on Acoustics, Speech, and Signal Processing},
  26:150--156, 1978.

\bibitem{lbj2}
Leland~B. Jackson.
\newblock An improved {Martinez/Parks} algorithm for {IIR} design with unequal
  numbers of poles and zeros.
\newblock {\em IEEE Transactions on Signal Processing}, 42(5):1234--1238, May
  1994.

\bibitem{saramaki}
T.~Saramaki.
\newblock Design of optimum recursive digital filters with zeros on the unit
  circle.
\newblock {\em IEEE Transactions on Acoustics, Speech, and Signal Processing},
  31:450--458, 1983.

\bibitem{alkhairy2}
Ashraf Alkhairy.
\newblock An efficient method for {IIR} filter design.
\newblock In {\em Proceedings of the IEEE International Conference on
  Acoustics, Speech, and Signal Processing}, volume~3, pages III:569--571, IEEE
  ICASSP-94, Adelaide, Australia, April 19--22 1994.

\bibitem{RGH}
L.~R. Rabiner, N.~Y. Graham, and H.~D. Helms.
\newblock Linear programming design of {IIR} digital filters with arbitrary
  magnitude function.
\newblock {\em IEEE Transactions on Acoustics, Speech, and Signal Processing},
  ASSP-22(2):117--123, April 1974.

\bibitem{limlee}
Y.~C. Lim, J.~H. Lee, C.~K. Chen, and R.~H. Yang.
\newblock A weighted least squares algorithm for quasi--equripple {FIR} and
  {IIR} digital filter design.
\newblock {\em IEEE Transactions on Signal Processing}, 40(3):551--558, March
  1992.

\bibitem{mclang7}
Mathias~C. Lang.
\newblock Least squares design of {IIR} filters with arbitrary magnitude and
  phase responses and specified stability margin.
\newblock In {\em Proceedings of EUSIPCO-98}, Rhodes, Greece, September 1998.
\newblock paper \#194.

\bibitem{mclang8}
Mathias~C. Lang.
\newblock Weighted least squares {IIR} filter design with arbitrary magnitude
  and phase responses and specified stability margin.
\newblock In {\em Proceedings of DFSP-98}, Victoria, B.C., June 1998.

\bibitem{lang11}
Mathias~C. Lang.
\newblock Least-squares design of {IIR} filters with prescribed magnitude and
  phase responses and a pole radius constraint.
\newblock {\em {IEEE} Transactions on Signal Processing}, 48(11):3109--3121,
  November 2000.

\bibitem{J&L}
L.~B. Jackson and G.~J. Lemay.
\newblock A simple {Remez} exchange algorithm for design {IIR} filters with
  zeros on the unit circle.
\newblock In {\em Proceedings of the ICASSP-90}, page 13.D3.15, Albuquerque,
  NM, April 1990.

\bibitem{S&A}
V.~L. Stonick and S.~T. Alexander.
\newblock Global optimal {IIR} filter design and {ARMA} estimation using
  homotopy continuation methods.
\newblock In {\em Proceedings of the ICASSP-90}, page 13.D3.13, Albuquerque,
  NM, April 1990.

\bibitem{cain}
A.~Tarcz{\'n}ski, B.~D. Cain, E.~Hermanowicz, and M.~Rojewski.
\newblock A {WISE} method for designing {IIR} filters.
\newblock {\em IEEE Transactions on Signal Processing}, 49(7):1421--1432, July
  2001.

\bibitem{graham}
L.~R. Rabiner, N.~Y. Graham, and H.~D. Helms.
\newblock Linear programming design of {IIR} digital filters with arbitrary
  magnitude function.
\newblock {\em IEEE Transactions on Acoustics, Speech, and Signal Processing},
  22:117--123, 1974.

\bibitem{crosara}
S.~Crosara and G.~A. Mian.
\newblock A note on the design of {IIR} filters by the differential-correction
  algorithm.
\newblock {\em IEEE Transactions on Circuits and Systems}, 30:898--903, 1983.

\bibitem{berchin}
Greg Berchin.
\newblock Precise filter design.
\newblock {\em {IEEE} Signal Processing Magazine}, 24(1):137--139, January
  2007.

\bibitem{steiglitz5}
K.~Steiglitz.
\newblock Computer-aided design of recursive digital filters.
\newblock {\em IEEE Transactions on Audio and Electroacoustics}, 18:123--129,
  1970.

\bibitem{deczky5}
A.~G. Deczky.
\newblock Synthesis of recursive digital filters using the minimum p-error
  criterion.
\newblock {\em IEEE Transactions on Audio and Electroacoustics}, 22:98--111,
  1974.

\bibitem{cadzow5}
J.~A. Cadzow.
\newblock Recursive digital filter synthesis via gradient based algorithms.
\newblock {\em IEEE Transactions on Acoustics, Speech, and Signal Processing},
  24:349--355, October 1976.

\bibitem{dudgeon2}
D.~E. Dudgeon.
\newblock Recursive filter design using differential correction.
\newblock {\em IEEE Transactions on Acoustics, Speech, and Signal Processing},
  22:443--448, 1974.

\bibitem{lbj3}
L.~B. Jackson.
\newblock Frequency-domain {Steiglitz-McBride} method for least-squares {IIR}
  filter design, {ARMA} modeling, and periodogram smoothing.
\newblock {\em {IEEE} Signal Processing Letters}, 15:49--52, 2008.

\end{thebibliography}

\end{document}